\documentclass[opre]{informs3}
\usepackage{endnotes}
\let\footnote=\endnote

\usepackage{enumerate}
\usepackage{verbatim}
\usepackage{amsmath,amssymb}
\usepackage{subfigure}
\usepackage{url}

\newtheorem{thm}{Theorem}

\newcommand{\beql}[1]{\begin{equation}\label{#1}}
\newcommand{\eeq}{\end{equation}}

\usepackage{graphicx}
\usepackage[english]{babel}
\usepackage{natbib}
 \bibpunct[, ]{(}{)}{,}{a}{}{,}%
 \def\bibfont{\small}%
 \def\newblock{\ }%

\TheoremsNumberedThrough

\ECRepeatTheorems

\EquationsNumberedThrough

\begin{document}

\setlength{\pdfpageheight}{\paperheight}
\setlength{\pdfpagewidth}{\paperwidth}

\RUNAUTHOR{Rama Cont \& Adrien de Larrard} \RUNTITLE{Price dynamics
in a Markovian limit order  market} \TITLE{Price dynamics in a
Markovian limit order  market} \ARTICLEAUTHORS{ \AUTHOR{Rama CONT \&
Adrien de LARRARD}
  \AFF{IEOR Dept, Columbia University, New York\\  \& \\ Laboratoire de Probabilit\'es et Mod\`eles Al\'eatoires\\ CNRS-Universit\'e Pierre et Marie Curie, Paris.}
    }

\ABSTRACT{ We propose and study a simple stochastic model for the dynamics of a limit order book,
in which arrivals of market order, limit orders and order cancellations are described in terms of
 a Markovian queueing system.
Through its analytical tractability, the model allows to obtain
analytical expressions for various quantities of interest such as
the distribution of the duration between  price changes, the
distribution and autocorrelation of price changes, and the
probability of an upward move in the price,  {\it conditional} on
the state of the order book. We study the diffusion limit of the
price process and express the volatility of price changes in terms
of parameters describing the arrival rates of buy and sell orders
and cancelations. These analytical results provide some insight into
the relation between order flow and price dynamics in order-driven
markets.} 

\maketitle

\tableofcontents\newpage
\clearpage
\section{Introduction}\label{secIntro}
An increasing number of stocks are traded in electronic, order-driven markets, in which
 orders to buy and sell are centralized in a {\it
  limit order book} available to to market participants and
market orders are executed against the best available offers in the limit order book.
The dynamics of prices in such markets are not only interesting from the viewpoint of market participants --for trading and order execution (\cite{schied10,shreve10})--
but also from a fundamental perspective, since they provide a rare glimpse into the dynamics of supply and demand and their role in
 price formation.

Equilibrium models of price formation in limit order markets
(\cite{parlour,rosu09}) have shown that the evolution of the  price
in such markets is rather complex and depends on the state of the
order book. On the other hand, empirical studies on  limit order
books
(\cite{bouchaud08,FarmerLarge,gourieroux99,hollifield1,smith2003})
provide an extensive list of statistical features of order book
dynamics that are challenging to incorporate in a single model.
While most of these studies have focused  on
unconditional/steady--state distributions of various features of the
order book, empirical studies (see e.g. \cite{harris05}) show that
the state of the order book contains information on short-term price
movements so it is of interest to provide forecasts of various
quantities {\it conditional } on the state of the order book.
Providing analytically tractable models which enable to compute
and/or reproduce {\it conditional} quantities which are relevant for
trading and intraday risk management has proven to be challenging,
given the complex relation between order book dynamics and price
behavior.

The search for tractable models of limit order markets has led to the development of stochastic models
which aim to retain the main statistical features of  limit order books while remaining computationally manageable.
 Stochastic models also serve to illustrate how far
one  can go in reproducing the dynamic properties of a limit order book
  without resorting to detailed behavioral assumptions about market
  participants or introducing unobservable parameters describing agent
  preferences, as in more detailed market microstructure models.

Starting from a description of order arrivals and cancelations as
point processes, the dynamics of a limit order book is naturally
described in the language of queueing theory. \cite{englelunde03}
formulates a bivariate point process to jointly analyze trade and
quote arrivals. \cite{cont2010} model the dynamics of a limit order
book as a tractable multiclass queueing system and compute various
transition probabilities of the price conditional on the state of
the order book, using Laplace transform methods.

\subsection{Summary}

We propose in this work a Markovian model of a limit order market,
which captures some salient features of the dynamics of market
orders and limit orders, yet
 is even simpler than the model of  \cite{cont2010} and enables
a wide range of properties of the price process to be computed
analytically.

Our approach is motivated by the observation that, if one is primarily interested in the dynamics of the price, it is sufficient to focus on the dynamics of
the (best) bid and ask queues.
Indeed, empirical evidence shows that  most of the order flow is directed at the best bid and ask prices (\cite{biais95}) and the imbalance between the order flow at the bid and at the ask appears to be the main driver of
price changes (\cite{cks2010}).

Motivated by this remark, we propose a parsimonious model in which
the limit order book is represented by the number of limit orders
$(q_{t}^{b},q_{t}^{a})$ sitting at the bid and the ask, represented
as a system of two interacting  queues. The remaining levels of the
order book are treated as a `reservoir' of limit orders represented
by the distribution of the size of the queues at the 'next-to-best'
price levels. Through its analytical tractability, the Markovian
version of our model allows to obtain analytical expressions for
various quantities of interest such as the distribution of the
duration until the next price change, the distribution and
autocorrelation of price changes, and the probability of an upward
move in the price, {\it conditional} on the state of the order book.

Compared with econometric models of high frequency data
\cite{engle98,englelunde03} where the link between durations and
price changes is specified exogenously, our model links these
quantities in an endogenous manner, and provides a first step
towards joint 'structural' modeling of high frequency dynamics of
prices and order flow.

A second important observation is that order arrivals and
cancelations are very frequent and occur at millisecond time scale,
whereas, in many applications such as order execution, the metric of
success is the volume-weighted average price (VWAP) so one is
interested in the dynamics of order flow over a large time scale,
typically tens of seconds or minutes. As shown in Table
\ref{heavytraffic.table}, thousands of order book events may occur
over such time scales. This {\it aggregation} of events actually
simplifies much of the analysis and enables us to use asymptotic
methods. We study the link between price volatility and order flow
in this model by studying the {\it diffusion limit} of the price
process. In particular, we express the volatility of price changes
in terms of parameters describing the arrival rates of buy and sell
orders and cancelations. These analytical results provide some
insight into the relation between order flow and price dynamics in
order-driven markets. Comparison of these asymptotic results with
empirical data shows that main insights of the model to be correct:
in particular, we show that in limit order markets where orders
arrive frequently, the volatility of price changes is increase with
the ratio of the order arrival intensity to market depth, as
predicted by our model.
\begin{table}[h]
\begin{center}
\hspace{0.5cm}
\begin{tabular}{|c|c|c|}
\hline
   & Average no. of & Price changes\\
    & orders in 10s &  in 1 day \\ \hline
Citigroup  & 4469 & 12499 \\ \hline

General Electric  & 2356 & 7862 \\ \hline

General Motors  & 1275 &  9016 \\ \hline
\end{tabular}
\vskip 0.5cm
\caption{Average number of orders in  10 seconds and number of price changes (June 26th,  2008).}
\end{center}
\label{heavytraffic.table}\end{table}

\subsection{Outline}
The paper is organized as follows.
Section \ref{orderbook.sec}  introduces  a reduced-form representation of a limit order book
and presents a Markovian model
in which limit orders, market orders and cancellations occur according to  Poisson processes.
Section \ref{analytical.sec} presents various analytical results for this model: we compute the distribution of the
duration until the next price change (section \ref{duration.sec}), the probability of upward move in the price (section \ref{upward.sec})
and the dynamics of the price (section \ref{pricedynamics.sec}).
In Section \ref{diffusionlimit.sec}, we show that the price behaves, at longer time scales,
 as a Brownian motion whose variance is expressed in terms of the parameters describing the order flow, thus establishing a link between volatility and
order flow statistics.
\section{A Markov model of limit order book dynamics}\label{orderbook.sec}


\subsection{Level-1 representation of a limit order book}
Empirical studies of limit order markets
suggest that the major component of the order flow occurs at the (best) bid and ask price levels (see e.g. \cite{biais95}). Furthermore, studies on
the price impact of order book events show that
the net effect of orders on the bid and ask queue sizes is the main factor driving price variations (\cite{cks2010}).
These observations, together with the fact that queue sizes at the best bid and ask (``Level I" order book) are
more easily obtainable (from trades and best quotes) than Level II data, motivate a
reduced-form modeling approach in which we represent the state of the limit order book
by
\begin{itemize}\item the bid price $s_t^b$ and the ask price $s_t^a$
\item the size of the bid queue  $q_t^b$ representing the outstanding limit buy orders at the bid, and
\item the size of the ask queue $q_t^a$  representing the outstanding limit sell orders at the ask
\end{itemize}
Figure 1 summarizes this representation.
\begin{figure}[tbh]
\begin{center}
\includegraphics[width=9cm]{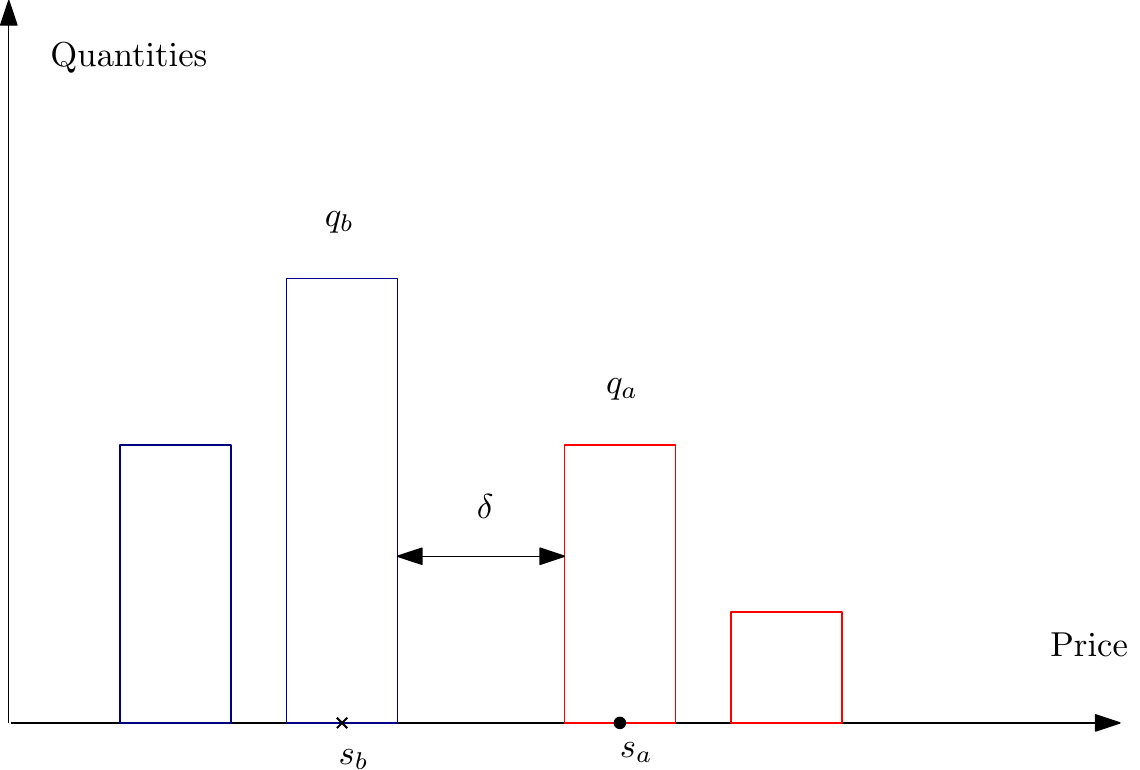}
\caption{Simplified representation of a  limit order book.}\end{center}
\label{orderbook.fig}
\end{figure}

The bid and ask prices are multiples of the tick size $\delta$.
As shown in Table \ref{spread.table}, for liquid stocks the  {\it bid-ask spread} $s_t^a-s^b_t$
is equal to one tick for more than $98\%$ of observations. We will therefore make the simplifying assumption
that the spread is equal to one tick, i.e. $s_t^a=s_t^b+\delta$, resulting in a further reduction of dimension in the model.
\begin{table}[h]
\begin{center}
\hspace{0.5cm}
\begin{tabular}{|l|c|c|c|}
\hline
Bid-ask spread &  1 tick &  2 tick & $\geq$ 3 tick \\ \hline

Citigroup & 98.82 & 1.18 & 0 \\ \hline

General Electric  & 98.80 & 1.18 & 0.02 \\ \hline

General Motors  & 98.71 & 1.15 & 0.14 \\ \hline
\end{tabular}
\caption{Percentage of observations with a given  bid-ask spread (June 26th, 2008).}
\end{center}\label{spread.table}
\end{table}

The  state of the limit order book is thus described  by the triplet
$X_t=(s_t^b,q_t^b,q_t^a)$ which takes values in   the discrete state space
$\delta. \mathbb{Z} \times  \mathbb{N}^{2}$.

\subsection{Order book dynamics}
The state $X_t$ of the order book  is modified by {\it order book events}:
limit orders (at the bid or ask), market orders and cancelations (see \cite{cont2010,cks2010,smith2003}).
A limit buy (resp. sell) order of size $x$ increases the size of the bid (resp. ask) queue by $x$, while
a market buy (resp. sell) order decreases the corresponding queue size by $x$.
Cancellation of $x$ orders in a given queue reduces the queue size by $x$.
Given that we are  interested in the queue sizes at the best bid/ask levels, market orders and
cancellations have the same effect on the state variable $X_t$.

\begin{figure}[tbh]     \centering     \includegraphics[width=0.7\textwidth]{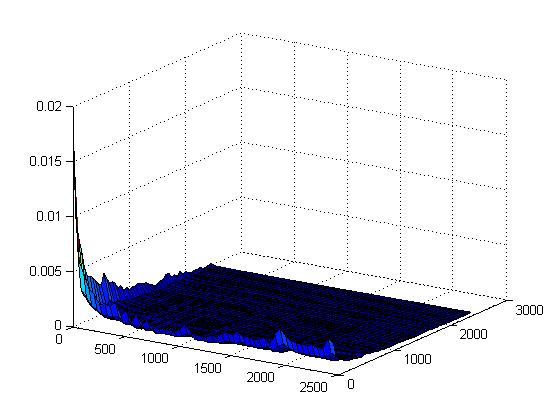}
\caption{Joint density of bid and ask queue sizes after a price move (Citigroup, June 26th 2008).}
\label{f.fig}
\end{figure}

We will assume that these events occur according to independent Poisson processes: 
\begin{itemize}
   \item Market buy (resp. sell)
   orders arrive at independent, exponential times with rate $\mu$,
   \item Limit buy (resp. sell) orders at the (best) bid (resp. ask) arrive at independent, exponential times with rate
   $\lambda$,
   \item Cancellations occur at independent, exponential times with rate $\theta$.
   \item These events are mutually independent.
   \item All orders sizes are equal (assumed to be 1 without loss of generality).
\end{itemize}
Denoting by $(T_{i}^{a},i\geq 1)$ (resp. $T_{i}^{b}$) the times at which the size of ask (resp.   the bid) queue changes  and $V_{i}^{a}$ (resp. $V_{i}^{a}$) the size of the associated change in queue size, the above assumptions translate into the following properties for the sequences $T_{i}^{a},T_{i}^{b},V_{i}^{a},V_{i}^{b}$:
\begin{itemize}
\item[(i)]   $(T_{i+1}^{a}-T_i^a)_{i \geq 0}$ is a sequence of independent random variables with exponential distribution with parameter $\lambda + \theta + \mu$,
\item[(ii)]  $(T_{i+1}^{b}-T_i^b)_{i \geq 0}$ is a sequence of independent random variables with exponential distribution with parameter $\lambda + \theta + \mu$,
\item[(iii)]   $(V_{i}^{a})_{i \geq 0}$ is a sequence of independent random variables with
 \begin{equation}\mathbb{P}[V_{i}^{a} = 1] = \dfrac{\lambda}{\lambda + \mu + \theta} \ \  and \ \  \mathbb{P}[V_{i}^{a} = -1] = \dfrac{\mu + \theta}{\lambda + \mu + \theta},\label{Via.eq}\end{equation}
\item[(iv)]   $(V_{i}^{b})_{i \geq 0}$ is a sequence of independent random variables with
\begin{equation}\mathbb{P}[V_{i}^{b} = 1] = \dfrac{\lambda}{\lambda + \mu + \theta} \ \  and \ \  \mathbb{P}[V_{i}^{b} = -1] = \dfrac{\mu + \theta}{\lambda + \mu + \theta}\label{Vib.eq}\end{equation}
\item All the previous sequences are independent.
\end{itemize}
Once the bid (resp. the ask) queue is depleted, the price will move to the queue at the next level,
which we assume to be one tick below (resp. above). The new queue size then corresponds to what was previously the number of orders sitting at the price immediately below (resp. above) the best bid (resp. ask).
Instead of keeping track of these queues (and the corresponding order flow) at all price levels (as in \cite{cont2010,smith2003}),
we treat these sizes as stationary variables drawn from a certain distribution $f$  on $\mathbb{N}^2$. Here $f(x,y)$  represents the probability of observing $(q^b_t,q^a_t)=(x,y)$ right after a price increase.
Similarly, we denote $\tilde{f}(x,y)$  the probability of observing $(q^b_t,q^a_t)=(x,y)$ right after a price decrease. More precisely, denoting by  ${\cal F}_t$ the history of prices and order book events on $[0,t]$,
\begin{itemize}
\item if $q^a_{t-}=0$ then $(q^b_t,q^a_t)$ is a random variable with distribution $f$, independent from ${\cal F}_{t-}$.
    \item if $q^b_{t-}=0$ then $(q^b_t,q^a_t)$ is a random variable with distribution $\tilde{f}$, independent from ${\cal F}_{t-}$.
\end{itemize}
Given the  independence assumptions on event types, the probability  that these two situations occur simultaneously is zero.
\begin{remark}
The asumption that $(q^b_t,q^a_t)$is independent from ${\cal F}_{t-}$ is not necessary. If one only assume that the random variables used to replace the quantity of orders once the price moves are stationnary, all the results from this paper remain valid. However, without this assumption, the process $(q^b_t,q^a_t)_{t \geq 0}$ becomes non-Markovian.
\end{remark}
The distributions $f$ and $\tilde{f}$ summarize the interaction of the queues at the best bid/ask levels with the rest of the order book,
viewed here as a 'reservoir' of limit orders.
For simplicity we shall assume $\tilde{f}(x,y)=f(y,x)$ i.e. events occurring on the bid and on the ask side have similar statistical properties but our analysis may be readily extended to the asymmetric case.
Figure \ref{f.fig} shows the (joint) empirical distribution of bid and ask queue sizes after a price move for Citigroup stock on
June 26th 2008.

Under these assumptions $q_t = (q^{b}_t, q^{a}_t)$ is thus a Markov process, taking values in $\mathbb{N}^2$, whose transitions correspond to the order book events $\{ T_i^a,i\geq 1\}\cup \{ T_i^b, i\geq 1\}$:
 \begin{itemize}
   \item At the arrival of a new limit  buy (resp. sell) order  the bid (resp. ask) queue increases by one unit. This occurs at rate $\lambda$.
   \item At each cancellation or market order, which occurs at rate $\theta+\mu$,
  either:\begin{itemize}
  \item[(a)] the corresponding queue decreases by one unit if it is $>1$, or
 \item[(b)] if the ask queue is depleted   then $q_t$  is a random variable with   distribution $f$.
     \item[(c)] if the bid queue is depleted      then $q_t$  is a random variable with   distribution $\tilde{f}$.
      \end{itemize}
 \end{itemize}

The values of $\lambda$ and $\mu+\theta$ are readily estimated from high-frequency records of order books (see \cite{cont2010} for a description of the estimation procedure). Table \ref{table.intensity} gives examples of such
 parameter estimates for the stocks mentioned above. We note that in all cases $\lambda < \mu+\theta$ but that the difference is
 small: $|(\mu+\theta)-\lambda|\ll \lambda$.
\begin{table}[h]\label{table.intensity}
\begin{center}
\hspace{0.5cm}
\begin{tabular}{|c|c|c|}
\hline
   & $\hat{\lambda}$  & $\hat{\mu}+\hat{\theta}$\\
   \hline
Citigroup  & 2204 & 2331 \\ \hline

General Electric  & 317 & 325 \\ \hline

General Motors  & 102 &  104 \\ \hline
\end{tabular}
\vskip 0.5cm
\caption{Estimates for the intensity of limit orders and market orders+cancellations, in number of batches per second (each batch representing 100 shares) on June 26th, 2008).}
\end{center}
\end{table}

 \subsection{Price dynamics}
 When the bid or ask queue is depleted, the price moves up or down  to the next level of the order book.
 We will assume that the order book contains no `gaps' (empty levels) so that these price increments are   equal to one tick:
 \begin{itemize}
   \item When  the bid queue is depleted, the price decreases by one tick.
   \item When  the ask queue is depleted, the price increases by one tick.
 \end{itemize}
If there are  gaps  in the order book, this results in 'jumps' (i.e. variations of more than one tick) in the price.
The price process $s^b_t$ is thus a piecewise constant process whose transitions correspond to hitting times of the  $\{ (0,y), y\in \mathbb{N}\}\cup \{(x,0), x\in \mathbb{N}\}$
by the Markov process $q_t = (q^{a}_t, q^{b}_t)$.
\subsection{Summary}
In summary, the process $X_t = (s^b_t,q^{b}_t,q^{a}_t)$ is   a  continuous-time process with right-continuous, piecewise constant  sample paths whose transitions correspond to the order book events $\{ T_i^a,i\geq 1\}\cup \{ T_i^b, i\geq 1\}$.
At each event:
\begin{itemize}
\item If an order or cancelation  arrives on the ask side i.e. $T\in \{ T_i^a,i\geq 1\}$:
$$(s_{T}^{b},q_{T}^{b},q_{T}^{a}) = (s_{T-}^{b},q_{T-}^{b},q_{T-}^{a} + V_{i}^{a}) 1_{q_{T-}^{a} > -V_{i}^{a}} + (S_{T-}^{b}+\delta,R_{i}^{b},R_{i}^{a}) 1_{q_{T-}^{a} \leq -V_{i}^{a}}, $$
\item If an order or cancelation  arrives on the bid side i.e. $T\in \{ T_i^b,i\geq 1\}$:
$$(s_{T}^{b},q_{T}^{b},q_{T}^{a}) = (s_{T-}^{b},q_{T-}^{b} + V_{i}^{b},q_{T-}^{a}) 1_{q_{T-}^{b} > -V_{i}^{b}} + (s_{T-}^{b}-\delta,\tilde{R}_{i}^{b},\tilde{R}_{i}^{a}) 1_{q_{T-}^{b} \leq -V_{i}^{b}}, $$
where $(V_{i}^{a})_{i\geq 1}$ and $(V_{i}^{b})_{i\geq 1}$ are sequences of IID variables with distribution given by \eqref{Via.eq}-\eqref{Vib.eq},
$(R_i)_{i\geq 1}=(R_{i}^{b},R_{i}^{a})_{i\geq 1}$ is a sequence of IID variables with (joint) distribution $f$, and
$(\tilde{R}_i)_{i\geq 1}=(\tilde{R}_{i}^{b},\tilde{R}_{i}^{a})_{i\geq 1}$ is a sequence of IID variables with (joint) distribution $\tilde{f}$.
\end{itemize}
\begin{remark}[Independence assumptions] The  IID assumption for the
sequences $(R_n),(\tilde{R}_n)$ is only used in Section
\ref{diffusionlimit.sec}. The results of Section
\ref{analytical.sec} do not depend on this assumption.
\end{remark}
\subsection{Quantities of interest}
In applications, one is interested in computing various quantities that intervene in high frequency trading such as:
\begin{itemize}
\item the conditional distribution of the duration between price moves, given the state of the order book (Section \ref{duration.sec}),
    \item the probability of a price increase, given the state of the order book (Section \ref{upward.sec}), \item the dynamics of the price :autocorrelations and distribution and autocorrelations of price changes (section \ref{pricedynamics.sec}), and
\item the volatility of the price (section \ref{diffusionlimit.sec}).
\end{itemize}
We will show that all these quantities may be characterized analytically in this model, in terms of order flow statistics.
\section{Analytical results}\label{analytical.sec}
The high-frequency dynamics of the price may be described in terms
of {\it durations} between successive price changes and the
magnitude of these price changes. Given that the state of the (Level
I) order book is observable, it is of interest to examine what
information  the current state of the order book gives about the
dynamics of the price. We now proceed to show how the model
presented above may be used to compute the conditional distributions
of durations and price changes, given the current state of the order
book, in terms of the arrival rates of market orders, limit orders
and cancellations. The result of this section do not depend on the
assumptions on the sequences $(R_n),(\tilde{R}_n)$.
\subsection{Duration until the next price change}\label{duration.sec}
We consider first the distribution of the duration until the next
price change, starting from a given configuration $(b,a)$ of the
order book. We define
\begin{itemize}
\renewcommand{\labelitemi}{$\bullet$}
\item $\sigma_{a}$ the first time when the  ask queue $(q_{t}^{a}, t \geq 0)$ is depleted,
\item $\sigma_{b}$ the first time when the bid queue $(q_{t}^{b}, t \geq 0)$ is depleted
\end{itemize}
Since the queue sizes are constant between events, one can express these stopping times as:
$$ \sigma^{a} = \inf \lbrace T_i^a, \ q_{T_i^a-}^{a}+V^a_i =  0 \rbrace
\qquad \sigma^{b} = \inf \lbrace T_i^b, \ q_{T_i^b-}^{b}+V^b_i = 0 \rbrace  $$
The price $(s_{t}, t \geq 0)$ moves when the queue $q_{t} = (q_{t}^{b},q_{t}^{a})$ hits one of the axes: the duration until the next price move is thus
$$ \tau = \sigma_{a} \wedge \sigma_{b}. $$
The following theorem gives the distribution of the duration $\tau$, conditional on the initial queue sizes:
\begin{proposition}[Distribution of duration until next price move] 
\label{laplaceHT}
The distribution of $\tau$ conditioned on the state of the order book is given by:
\begin{equation}\mathbb{P}[\tau >t| q_{0}^{a} = a, \ q_{0}^{b} = b] =
\sqrt{(\dfrac{\mu + \theta}{\lambda})^{a+b}}  \psi_{a,\lambda,\theta+\mu}(t) \psi_{b,\lambda,\theta+\mu}(t)\label{duration.eq}
\end{equation}
\begin{equation}{\rm where}\qquad \psi_{n,\lambda,\theta+\mu}(t)=\int_{t}^{\infty}  \dfrac{n}{u} I_{n}(2 \sqrt{\lambda (\theta+\mu)} u) e^{-u (\lambda +  \theta+\mu)} du\label{psi.eq}
\end{equation}
and $I_{n}$ is the modified Bessel function of the first kind.
  The conditional law of $\tau$ has a  regularly varying tail
\begin{itemize}\item with tail exponent $2$ if $\lambda <  \mu + \theta$
\item with tail exponent 1 if $\lambda =  \mu + \theta$. In particular, if  $\lambda =  \mu + \theta$,
$E[\tau  | q_{0}^{a}=a, q_{0}^{b}=b]=\infty$ whenever ${a}>0, {b}>0$.
\end{itemize}
\end{proposition}
\begin{figure}[tbh]
\begin{center}
\includegraphics[width=0.8\textwidth]{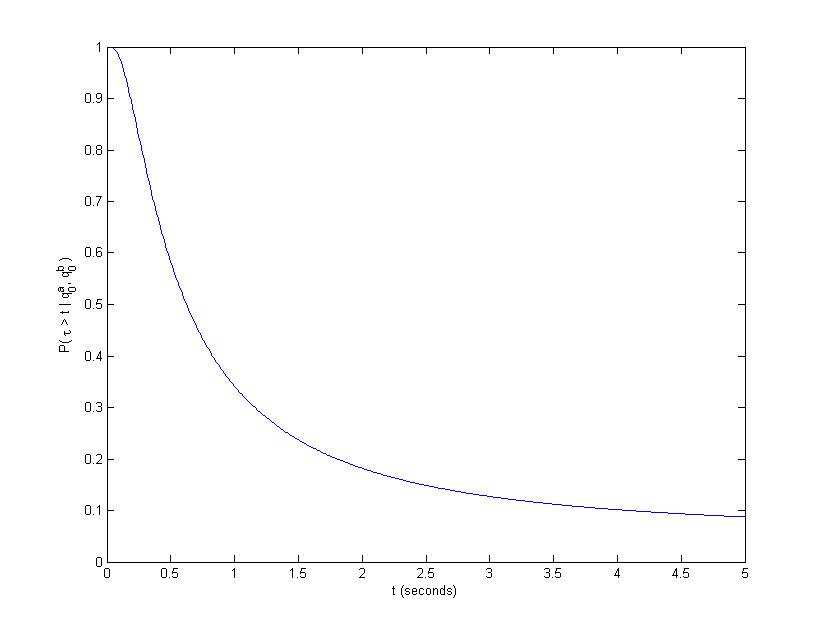}
\includegraphics[width=0.8\textwidth]{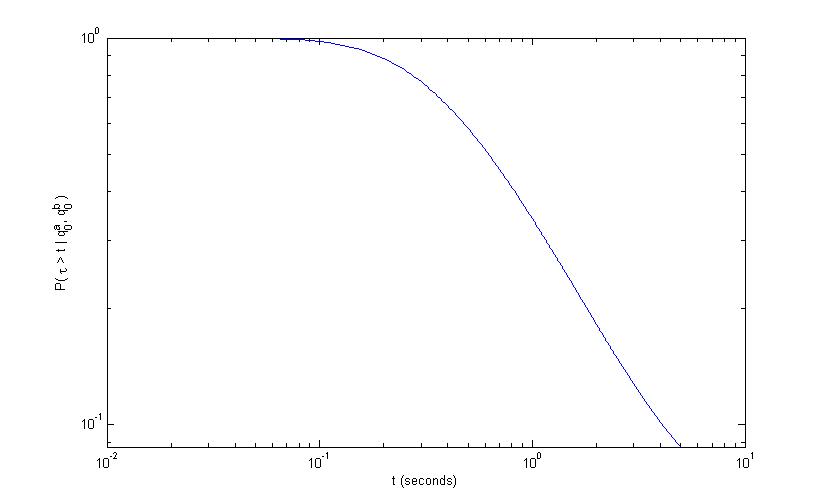}
\caption{Above: $P(\tau>t| q^a_0=4,q^b_0=5)$ as a function of $t$
for $\lambda=12, \mu+\theta=13$. Below: same figure in log-log
coordinates. Note the Pareto tail which decays as
$t^{-2}$.}\end{center} \label{duration.fig}
\end{figure}
\proof{Proof.}
Since $(q_{t}^{a}, t \geq 0)$ follows a birth and death process with birth rate $\lambda$ and death rate $\mu + \theta$, $\mathcal{L}(s,x) := \mathbb{E}[e^{-s\sigma_{a}}| q_{0}^{a} = x]$ satisfies:
$$ \mathcal{L}(s,x) = \dfrac{\lambda \mathcal{L}(s,x+1) + (\mu + \theta) \mathcal{L}(s, x-1)}{\lambda + \mu + \theta + s} .$$
We can find the roots of the polynomial: $\lambda X^{2} - (\lambda + \mu + \theta + s) X + \mu + \theta$; one root is $>1$, the other is $<1$; since $\mathcal{L}(s,0) = 1$ and $\lim_{x \rightarrow \infty} \mathcal{L}(s,x) = 0$,
$$ \mathcal{L}(s,x) = (\dfrac{(\lambda + \mu + \theta + s) - \sqrt{((\lambda + \mu + \theta + s))^{2} - 4\lambda (\mu + \theta)}}{2 \lambda})^{x}. $$

Moreover if we use the relation $\mathbb{P}[\tau >t| q_{0}^{a} = x, q_{0}^{b} = y] =\mathbb{P}[\sigma^{a} >t| q_{0}^{a}  = x]\mathbb{P}[\sigma^{b} >t| q_{0}^{b} = y]$,
$$\mathbb{P}[\tau >t| q_{0}^{a} = x, q_{0}^{b} = y] = \int_{t}^{\infty} \hat{\mathcal{L}}(u,x) du \int_{t}^{\infty} \hat{\mathcal{L}}(u,y) du .$$
This Laplace transform may be inverted (see  \cite[XIV.7]{Feller1971})
and the inversion yields
 $$\hat{\mathcal{L}}(t,x)  = \dfrac{x}{t} \sqrt{(\dfrac{\mu + \theta}{\lambda})^{x}}\quad I_{x}(2 \sqrt{\lambda(\theta + \mu)} t) e^{-t (\lambda + \theta + \mu)}, $$
which gives us the expected result.
\endproof

\textbf{Tail behavior of $\tau$:}
\begin{itemize}
\renewcommand{\labelitemi}{$\bullet$}
\item If $\lambda < \mu + \theta$: $$ \mathcal{L}(s,x) = \alpha(s)^{x} \mathop{\sim}_{ s \rightarrow 0} 1 - \dfrac{x(\lambda+ \mu+ \theta)}{2 \lambda(\mu + \theta - \lambda)} s,  $$
so Karamata's Tauberian theorem  \cite[XIII.5]{Feller1971} yields
$$ \mathbb{P}[\sigma^{a} > t|q_{0}^{a} = x]  \mathop{\sim}_{ t \rightarrow \infty} \dfrac{x(\lambda+ \mu+ \theta)}{2 \lambda(\mu + \theta- \lambda)} \dfrac{1}{t};  $$
therefore the conditional law of the duration $\tau$ is
a regularly varying  with tail index $2$
\begin{equation}\mathbb{P}[\tau > t|q_{0}^{a} = x, q_{0}^{b} = y]  \mathop{\sim}_{ t \rightarrow \infty}  \dfrac{xy(\lambda+ \mu+ \theta)^{2}}{ \lambda^{2}(\mu + \theta - \lambda)^{2}}  \dfrac{1}{4t^2}.  \label{tautail.dom}\end{equation}
\item If the order flow is balanced i.e. $\lambda = \mu + \theta$ then
$$ \mathcal{L}(s,x) =(\alpha(s))^{x} \mathop{\sim}_{ s \rightarrow 0} 1 - \dfrac{x}{\sqrt{\lambda}} \sqrt{s},$$
the law of $\sigma^{a}$ is regularly-varying with tail index $1/2$ and
$$ \mathbb{P}[\sigma^{a} > t | q_{0}^{a} = x]  \mathop{\sim}_{ t \rightarrow \infty} \dfrac{x}{\sqrt{\pi \lambda}} \dfrac{1}{\sqrt{t}}. $$
The duration then follows a heavy-tailed distribution with infinite first moment:
\begin{equation} \mathbb{P}[\tau > t| q_{0}^{a} = x, q_{0}^{b} = y]  \mathop{\sim}_{ t \rightarrow \infty} \dfrac{xy}{\pi \lambda} \dfrac{1}{t}; \label{tautail.balanced}\end{equation}
\end{itemize}












The expression given in \eqref{duration.eq} is easily computed  by discretizing the integral in \eqref{psi.eq}. Plotting \eqref{duration.eq}
for a fine grid of values of $t$ typically takes less than a second on a laptop.
Figure 3 
gives a numerical example, with $\lambda=12\ {\rm sec}^{-1}, \mu+\theta=13\ {\rm sec}^{-1}, q^a_0=4,q^b_0=5$ (queue sizes are given in multiples of average batch size).

\subsection{Probability of upward move in the price for a balanced limit order book}\label{upward.sec}
 Assume now that $\lambda = \mu + \theta$, i.e. that the flow of limit orders is balanced by the flow of market orders and cancellations.
 Therefore for all $t \leq \tau $, $q_{t} = M_{N_{2\lambda t}}$, where $(M_{n}, n \geq 0)$ is a symmetric random walk on $\mathbb{Z}^{2}$ killed when it hits either the x-axis or the y-axis and $(N_{2\lambda t}, t \geq 0)$ is a Poisson process with parameter $2\lambda$. Hence the probability of an upward move in the price  starting from a configuration $q^b_t=n,q^a_t=p$
 for the order book is equal to the probability that the random walk $M$ starting from $(n,p)$  hits the x-axis before the y-axis. This probability is given by the following proposition:
\begin{proposition}
For   $(n,p) \in \mathbb{N}^{2}$, the probability $\phi(n,p)$ that the next price move is an increase, conditioned on having the $n$
orders on the bid side and $p$ orders on the ask side is:
 \begin{equation} \phi(n,p) = \dfrac{1}{\pi} \int_{0}^{\pi} (2-\cos(t)-\sqrt{(2-\cos(t))^2-1})^{p}  \frac{\sin(nt) \cos(\frac{t}{2})}{\sin(\frac{t}{2})} dt.\label{phinp.eq}\end{equation}
\end{proposition}
\proof{Proof.}
 The generator of the bivariate random walk $(M_{n}, n \geq 1)$ is the discrete Laplacian so
 $\phi(n,p) = \mathbb{P}[\sigma_{a} < \sigma_{b}|q_{0-}^{b} = n, q_{0-}^{a} = p]$ satisfies, for all $n \geq 1$ and $p \geq 1$,
\begin{equation} 4\phi(n,p) = \phi(n+1,p) + \phi(n-1,p) + \phi(n,p+1) + \phi(n,p-1), \label{discreteDirichlet.eq}\end{equation}
with the boundary conditions: $\phi(0,p) = 0$ for all $p \geq 1$ and $\phi(n, 0 )= 1$ for all $n \geq 1$. This problem is known as the {\it discrete Dirichlet} problem;
solutions of \eqref{discreteDirichlet.eq} are called discrete harmonic functions. \cite[Ch. 8]{lawlervlada2010} show that for all $t \geq 0$, the functions
$$ f_{t}(x,y) = e^{xr(t)} \sin(yt),  \qquad {\rm and}  \qquad\tilde{f}_{t}(x,y) = e^{-xr(t)} \sin(yt)\qquad{\rm with}\quad r(t) = \cosh^{-1}(2 - \cos t) $$
are solutions of \eqref{discreteDirichlet.eq}.   In \cite[Corollary 8.1.8]{lawlervlada2010}
it is shown that the probability that a simple random walk $(M_{k}, k \geq 1)$ starting at $(n,p)\in \mathbb{Z}^+\times \mathbb{Z}^+$ reaches the axes at  $(x,0)$  is
$$ \dfrac{2}{\pi} \int_{0}^{\pi} e^{-r(t) p} \sin(nt) \sin(t x) dt,   $$
therefore
$$\phi(n,p) = \sum_{k=1}^{\infty} \dfrac{2}{\pi} \int_{0}^{\pi} e^{-r(t) p} \sin(t n) \sin(t k) dt . $$
Since
$$ \sum_{k=1}^{m} \sin(kt) = \dfrac{\sin(\frac{mt}{2})\sin(\frac{(m+1)t}{2})}{\sin(t/2)}=\dfrac{\cos(\frac{t}{2}) - \cos((m+\frac{1}{2})t)}{2 \sin(t/2)}, $$
using integration by parts we see that the second term leads to the integral:
$$ \int_{0}^{\pi}  \underbrace{\frac{e^{-r(t)p}\sin(nt)}{\sin(t/2)}}_{g(t)}\cos((m+1/2)t) dt= - \frac{1}{m+\frac{1}{2}}\int_{0}^{\pi} g'(t) \sin((m+\frac{1}{2})t) dt \mathop{\to}_{m\to\infty} 0.$$
since $g'$ is bounded. So finally:
$$\phi(n,p) =  \dfrac{1}{\pi} \int_{0}^{\pi} e^{-r(t) p} \sin(t n) \frac{\cos(\frac{t}{2})}{\sin(\frac{t}{2})} dt . $$
Noting that $e^{-r(t)}=(2-\cos(t)-\sqrt{(2-\cos(t))^2-1})$ we obtain the result.
\endproof

\begin{figure}[tbh]
\begin{center}
\includegraphics[width=0.8\textwidth]{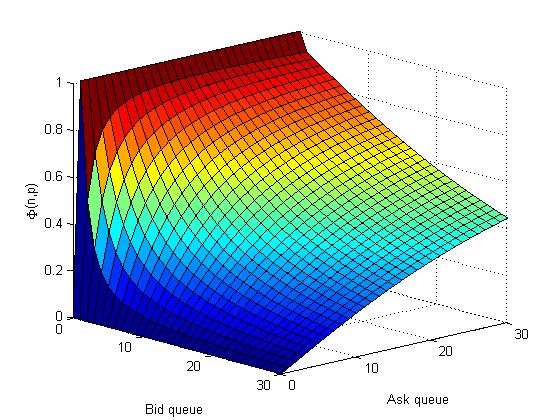}
\includegraphics[width=0.8\textwidth]{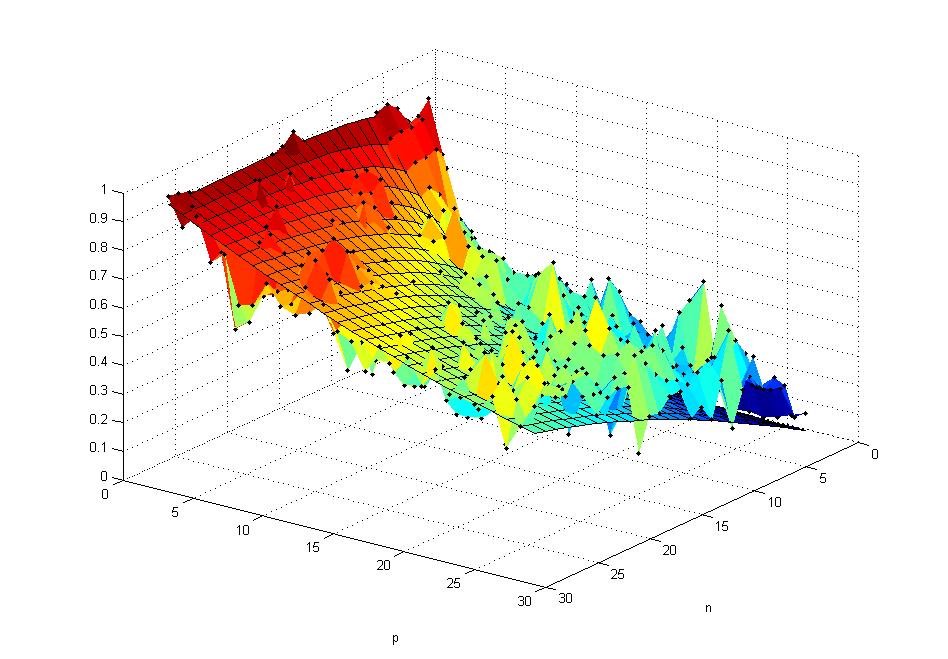}
\caption{Above: Conditional probability of a price increase, as a
function of the bid and ask queue size. Below: comparison with
transition frequencies for CitiGroup tick-by-tick data on June 26,
2008.}\end{center} \label{phinp.fig}\end{figure} Note that the
conditional probabilities \eqref{phinp.eq} are, in the case of a
balanced order book, independent of the parameters describing the
order flow.

The expression \eqref{phinp.eq} is easily computed numerically:
Figure 4 displays the shape of the function $\Phi$. The comparison
with the corresponding empirical transition frequencies for
CitiGroup tick-by-tick data on June 26, 2008  shows good agreement
between the theoretical conditional probabilities and their
empirical counterparts.

\subsection{Dynamics of the price}\label{pricedynamics.sec}
The high-frequency dynamics of the price in this model is described by a piecewise constant, right continuous process
 $(s_{t}, t \geq 0)$ whose jumps times correspond to times when the order book process $(q_{t}, t \geq 0)$ hits one of the axes. Denote by $(\tau_{1}, \tau_{2}, ...)$ the successive durations between price changes. The number of price changes  that occur during $[0,t]$ is given by
     $$ N_{t} := \max \lbrace \  n \geq 0, \ \  \tau_{1} + ...+ \tau_{n} \leq t \  \rbrace $$
At $t = \tau_{i}$, $s_{\tau_{i}} = s_{\tau_{i-}} + 1$ if $q_{\tau_{i-}^{a}} = 0$ and $s_{\tau_{i}} = s_{\tau_{i-}} - 1$ if $q_{\tau_{i-}^{b}} = 0$. $(X_{1}, X_{2}, X_{3},...,X_{n},...)$ are the successive moves in the price. Note that in general this is not a sequence of independent random variables. We define for $n \geq 1$,
 $$Z_{n} = \sum_{i=1}^{n} X_{i}$$
the value of the price, after $n$ changes. Hence, for all $t \geq 0$, $ s_{t} = Z_{N_{t}}. $
\begin{proposition}
Let $p_{cont} = \mathbb{P}[X_{2} = \delta | X_{1} = \delta] = \mathbb{P}[X_{2} = - \delta | X_{1} = - \delta] $
be the probability of two successive price moves in the same direction.
\begin{itemize}
\item $\forall k \geq 1, \ \ Cov(X_{1},X_{k}) = (2p_{cont}-1)^{k-1}. $
\item Conditional on the current state of the limit order book,  the distribution of the n-th subsequent price change $X_{n}$  is:
 $$p_{n}(x,y) := \mathbb{P}[X_{n} = \delta| q^{a}_{0} = x, \ q^{b}_{0} = y] = \dfrac{1+(2p_{cont}-1)^{n-1}(2p_{1}(x,y)-1)}{2}, $$
\end{itemize}
\end{proposition}
\proof{Proof.}
Let, for $(x,y) \in \mathbb{N}^{2}$, and for all $n \geq 2$, $p_{n}(x,y)$  the probability that $X_{n} = \delta$, conditioned on $ q^{a}_{0} = x$ and $q^{b}_{0} = y$. To simplify, we note $p_{n}$ for $p_{n}(x,y)$. $p_{n}$ is characterized by the following recurrence relation:
$$ \begin{pmatrix} p_{n} \\ 1-p_{n} \end{pmatrix} =   \begin{pmatrix}
p_{cont}&1-p_{cont} \\
1-p_{cont}&p_{cont}
\end{pmatrix} \begin{pmatrix} p_{n-1} \\ 1-p_{n-1} \end{pmatrix}, $$
hence
$$ \begin{pmatrix} p_{n} \\ 1-p_{n} \end{pmatrix} =   \begin{pmatrix}
p_{cont}&1-p_{cont} \\
1-p_{cont}&p_{cont}
\end{pmatrix} ^{n-1} \begin{pmatrix} p_{1} \\ 1-p_{1} \end{pmatrix}. $$

The eigenvalues of this matrix are $1$ and $2 p_{cont} -1$:

$$\begin{pmatrix}
p_{cont}&1-p_{cont} \\
1-p_{cont}&p_{cont}
\end{pmatrix}  =  \begin{pmatrix}
1 & 1 \\
1 &-1
\end{pmatrix}  \begin{pmatrix}
1 & 0 \\
0 & 2 p_{cont}-1
\end{pmatrix}  \begin{pmatrix}
1/2 & 1/2 \\
1/2 &-1/2
\end{pmatrix}.$$
Therefore
$$ p_{n} = \dfrac{1 + (2p_{cont} -1)^{n-1} (2 p_{1} -1)}{2} . $$
Moreover for all $n \geq 2$,
$$ Cov(X_{1},X_{n}) = p_{1}p_{n} + (1-p_{n})(1-p_{1}) -p_{1}(1-p_{n}) - p_{n}(1-p_{1}) $$
$$ Cov(X_{1},X_{n}) = (1 + 2p_{n}p_{1} - p_{n} - p_{1}) $$
$$ Cov(X_{1},X_{n}) = (2p_{cont}-1)^{n-1}. $$
\endproof
\begin{remark}[Negative autocorrelation of price changes at first lag] It is empirically observed that high frequency price movements have a negative autocorrelation at the first lag \cite{empirical}.
 In our model $Cov(X_{k},X_{k+1})< 0$ if and only if $p_{cont} < 1/2$, which happens when
$$ \sum_{i=1}^{\infty} \sum_{j\geq i}  f(i,j) > 1/2 $$
where $f$ is the joint distribution of queue sizes after a price increase.
This condition is verified on all high-frequency data sets we have examined. For example, for CitiGroup stock we find
$$ \sum_{i=1}^{\infty} \sum_{j\geq i}  f(i,j) > 0.7 $$
This asymmetry condition  on $f$ corresponds to the fact that, after an upward price  move,
 the new bid queue is generally smaller than the ask queue since the ask queue corresponds to the limit order
 previously sitting at second best ask level, while the bid queue results from the accumulation of orders over the very short period since the last price move.
Under this condition, high frequency  increments of the price are negatively correlated: an increase in the price is more likely to be followed by a decrease in the price.
\end{remark}
\begin{remark}
 The sequence of price increments  $(X_{1}, X_{2}, ...)$ is uncorrelated
if and only if $p_{cont} = 1/2$ which happens when
$$  \sum_{i=1}^{\infty} \sum_{j\geq i} f(i,j) = 1/2. $$
\end{remark}
\section{Diffusion limit of the price process} \label{diffusionlimit.sec}

As discussed in Section \ref{pricedynamics.sec}, the high frequency
dynamics of the price  is described by a piecewise constant
stochastic process $ s_{t} = Z_{N_{t}} $ where
$$ Z_{n} = X_{1} + ... + X_{n}  \ \ \ and \ \ N_{t} = \sup \lbrace k; \ \tau_{1} + ... + \tau_{k} \leq t \rbrace  $$
is the number of price moves during $[0,t]$.

However, over  time scales much larger than the interval between individual order book events,  prices are observed to have diffusive dynamics and modeled as such. To establish the link between the high frequency dynamics and the diffusive behavior at longer time scales, we shall  consider a time scale $t_{n} = t \zeta(n)$ over which the average number of order book events is of order $n$ and exhibit conditions under which the rescaled price process
$$ (s^{n}_{t} :=\dfrac{s_{t_{n}}}{\sqrt{n}}, t \geq 0)_{n \geq 1}  $$
verifies a functional central limit theorem i.e. converges in distribution to a non-degenerate process $(p_{t}, t \geq 0)$ as $n\to\infty$. The choice of the time scale $t_{n} = t \zeta(n)$ cannot be arbitrary: it is imposed by the distributional properties of the durations which, as observed in Section \ref{duration.sec}, are heavy tailed. More precisely, $\zeta(n)$ is chosen such that
$$\frac{\tau_1+...+\tau_n}{\zeta(n)}$$ has a well-defined limit.
In this section, we show that, under a symmetry condition, this
limit can be identified as a diffusion process whose diffusion
coefficient may be computed from the statistics of the order flow
driving the limit order book.

Assume $\lambda + \theta \leq \mu$ and  that the joint distribution
$f$ of the queue sizes after a price move satisfies:
\begin{eqnarray} D(f) = \sum_{i=1}^{\infty} \sum_{j=1}^{\infty} ij f(i,j)  < \infty \label{D(f).eq}\end{eqnarray}
The quantity $D(f)$ represents a measure of market depth: more
precisely, $\sqrt{D(F)}$ is the geometric average of the size of the
bid queue and the size of the ask queue after a price change.

In this section we  assume that the distribution $f$ is symmetric
with respect to its arguments: $\forall i,j \geq 0, \ f(i,j) =
f(j,i)$. Under this assumption, the sequence of increments $(X_{i},
i \geq 0)$ of the price is a sequence of independent random
variables. We will show that the limit $p$ is then a diffusion
process  which describes the dynamics of the price at lower
frequencies. In particular, we will compute the volatility of this
{\it diffusion limit} $p$ and relate it to the properties of the
order flow.

 In the following $\mathcal{D}$ denotes the space of right continuous  paths $\omega: [0, \infty) \rightarrow \mathbb{R}^{2}$ with left limits, equipped with the Skorokhod topology $J_{1}$, and $\Rightarrow$ will designate weak convergence on
 $(\mathcal{D},J_1)$ (see \cite{Billingsley,Whitt} for a discussion).
\subsection{Balanced order book} \label{sym}
We first consider the case of a {\it balanced} order flow for which the intensity of market orders and cancelations is equal to the intensity of limit orders. The study of high-frequency quote data indicates that this is an empirically relevant case for  many liquid stocks.
\begin{thm}\label{FCLT.balanced}
If $\lambda = \mu + \theta$,
$$ \left( \dfrac{s_{n\log n t}}{\sqrt{n}}, t \geq 0 \right) \mathop{\Rightarrow}^{n \to \infty} \left( \delta \sqrt{\dfrac{ \pi \lambda}{ D(f)}}  W_{t}, t \geq 0 \right)$$
where $\delta$ is the tick size, $D(f)$ is given by \eqref{D(f).eq}
and $W$ is a standard Brownian motion.
\end{thm}

\proof{Proof.}
For all $t \geq 0$ and $n \geq 1$, let $t_{n} = n \log n t$ and
\begin{equation}
 \dfrac{s_{n\log n t}}{\sqrt{n}} = \dfrac{ Z(t \pi \lambda /D(f))  \delta}{\sqrt{n}} +  \left(  \dfrac{Z(N_{t_
 {n}})\delta}{\sqrt{n}} -  \dfrac{ Z(t \pi \lambda /D(f))  \delta}{\sqrt{n}} \right)
\end{equation}
Using  Donsker's invariance principle, the sequence of processes
$(\dfrac{Z(t \pi \lambda n / D(f))}{\sqrt{n}}, t \geq 0)$ converges
in $({\cal D}, J_1)$ to a Brownian motion with volatility $ \delta
\sqrt{ \dfrac{ \pi \lambda}{ D(f)} }$.
 Let $\rho:(1,\infty)\mapsto (1,\infty)$ be a function satisfying:
$$ \rho(t) \log (\rho(t)) = t $$
Since $\rho(t) \mathop{\sim}_{ t \rightarrow \infty} \dfrac{t }{\log(t)}$, 
 \begin{equation}
 N_{t_n}  \mathop{\sim}_{ n \rightarrow \infty}  \rho(\dfrac{\pi \lambda t \zeta(n) }{D(f)})
  \sim  \dfrac{t \pi \lambda \zeta(n)}{D(f) \log(\zeta(n))  },
\end{equation}
 $$ N_{t_n} \mathop{\sim}_{ n \rightarrow \infty} \dfrac{ t \pi \lambda}{D(f)}.  $$
Therefore for all $t \geq 0$,
\begin{equation}
 \left(  \dfrac{Z(N_{t_{n}})\delta}{\sqrt{n}} -  \dfrac{ Z(t \pi \lambda /D(f))  \delta}{\sqrt{n}} \right) \mathop{\Rightarrow}^{n \to \infty} 0
\end{equation}
Therefore the finite dimensional distributions of the sequence of
processes $\left(  \dfrac{Z(N_{t_{n}})\delta}{\sqrt{n}} -  \dfrac{
Z(t \pi \lambda /D(f))  \delta}{\sqrt{n}} \right)_{ t \geq 0} $
converge to a point mass at zero . Since this sequence of processes
is tight on $(\mathcal{D},J_{1})$, it converges weakly to zero on
$(\mathcal{D},J_{1})$ (see \cite{Whitt}). Finally, $$ \left(
\dfrac{s_{n\log n t}}{\sqrt{n}}, t \geq 0 \right)
\mathop{\Rightarrow}^{n \to \infty}  \delta \sqrt{ \dfrac{ \pi
\lambda}{ D(f)} } W. $$
\endproof

\subsection{Empirical test using high-frequency data}
Theorem \ref{FCLT.balanced} relates the 'coarse-grained' volatility
of intraday returns at lower frequencies to the high-frequency
arrival rates of orders. Denote by $\tau_0= 1/\lambda$ the typical
time scale separating order book events. Typically $\tau_0$ is of
the order of milliseconds. In plain terms, Theorem
\ref{FCLT.balanced} states that, observed over a time scale
$\tau_{2} >> \tau_{0}$ (say, 10 minutes), the price has a diffusive
behavior with a diffusion coefficient given by
\begin{equation} \sigma =   \delta \sqrt{\frac{n  \pi \lambda}{ D(f)}} \label{volformula.eq}\end{equation}
 where $\delta$ is the tick size, $n$ is an integer verifying  $n\ln n \quad \tau_0= \tau_2$ which represents the average number of orders during an interval $\tau_2$
 and $\sqrt{D(F)}$, the geometric average of the size of the
bid queue and the size of the ask queue after a price change, is a
measure of market depth.

Formula \eqref{volformula.eq} links properties of the price to the
properties of the order flow. the left hand side represents the
variance of price changes, whereas the right hand side only involves
the tick size and quantities: it yields  an estimator for price
volatility which may be computed {\it without observing} the price!

The relation \eqref{volformula.eq} has an intuitive interpretation.
It shows that, in two 'balanced' limit order markets with the same
tick size and same rate of arrival of orders at the bext bid/ask,
the market with higher depth of the next-to-best queues will lead to
lower price volatility.

 More precisely, this formula shows that the microstructure of order flow affects price volatility through the ratio $\lambda/D(f)$ where $\lambda$ is the rate of arrival of limit orders and $D(f)$, given by \eqref{D(f).eq}, is a measure of market depth: in fact, our model predicts a proportionality between the variance of price increments and this ratio.  This is an empirically testable prediction: Figure
 5
 compares, for stocks in the Dow Jones index, the standard deviation of 10-minute price increments with  $\sqrt{\lambda/D(f)}$.

  We observe that, indeed, stocks with a higher  value of the ratio ${\lambda/D(f)}$ have a higher variance, and standard deviation of price increments increases roughly proportionally to $\sqrt{\lambda/D(f)}$.
 \begin{figure}[tbh]
\begin{center}
\includegraphics[width=0.7\textwidth]{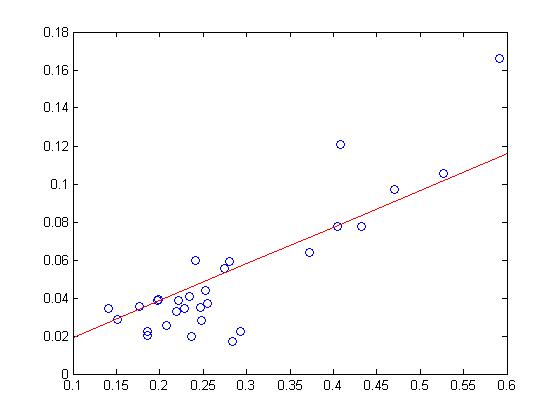}
\caption{$\sqrt{\lambda/D(f)}$, estimated from tick-by-tick order
flow (vertical axis) vs standard deviation of 10-minute price
increments (horizontal axis) for stocks in the Dow Jones Index,
estimated from high frequency data on June 26, 2008. Each point
represents one stock. Red line indicates the best linear
approximation.}\end{center} \label{empirical.fig}
\end{figure}
\subsection{Case when market orders and cancelations dominate} \label{asym}
We now consider the case in which the  flow of market orders and cancellations dominates that of limit orders: $\lambda < \theta + \mu$.
  In this case, price changes are more frequent since the order queues are depleted at a faster rate than they are replenished by market orders. We also obtain a diffusion limit though with a different scaling:
\begin{thm}\label{transientdiffusion.thm}
Let $\lambda < \theta + \mu$ and $f$ a probability distribution on $\mathbb{N}^{2}$ which satisfies
 $$ m(\lambda,\theta + \mu,f) = \sum_{i = 1}^{\infty} \sum_{j=1}^{\infty} m(\lambda,\theta + \mu,i ,j) f(i,j) < \infty, $$
where for all $(x,y) \in (\mathbb{N}^{*})^{2}$,
 $$ m(\lambda,\theta + \mu,x,y) = \int_{0}^{\infty} dt \int_{t}^{\infty} \psi_{x,\lambda,\mu+\theta}(u) du \int_{t}^{\infty} \psi_{y,\lambda,\mu+\theta}(u) du  $$
where $\psi_{x,\lambda,\mu+\theta}$ is given by \eqref{psi.eq}. Then
$$ \left( \dfrac{s_{nt}}{\sqrt{n}}, t \geq 0 \right) \mathop{\Rightarrow}^{n \to \infty} \left(  \sqrt{ \dfrac{1}{m(\lambda, \theta + \mu,f)} }\delta W_{t}, t \geq 0 \right)  $$
where $W$ is a standard Brownian motion.
\end{thm}
\proof{Proof.}
The sequence $(\tau_{2},\tau_{3},...)$ is a sequence of i.i.d random variables with finite mean equal to $m(\lambda,\theta + \mu,f)$. We apply the law of large numbers:
$$ \dfrac{ \tau_{1} + \tau_{2} + ... + \tau_{n}}{n} \mathop{\to}^{n \rightarrow \infty} m(\lambda,\theta+\mu,f). $$
Therefore,
$$ \forall t \geq 0,\qquad N^{n}_{t} \mathop{\sim}^{n \to \infty} [\dfrac{t n}{m(\lambda, \theta+\mu,f)}]. $$
The rest of the proof follows the lines of  the proof of theorem \ref{FCLT.balanced}.
\endproof

\paragraph{Variance of  price change at intermediate frequency}
Similarly to Theorem \ref{FCLT.balanced}, Theorem
\ref{transientdiffusion.thm} leads to an expression of the variance
of the price at a time scale $\tau
>> \tau_{0}$, where $\tau_{0}(\sim$ ms) is the average interval
between order book events:
 \begin{equation} \sigma^{2} =  \frac{\tau}{\tau_0} \dfrac{\pi \lambda}{  m(\lambda, \theta+\mu,f) } \delta^{2}\end{equation}
Here, $ m(\lambda, \theta+\mu,f) $ represents the expected hitting
time of the axes by the queueing system with parameters $(\lambda,
\theta+\mu)$ and random initial condition with distribution $f$ in
the positive orthant.

 As before, while the left hand side of this equation is the variance of price changes (over a time scale  $\tau_2$), the right hand side only involves the tick size and
 quantities which relate to the statistical properties of the order flow.
\subsection{Conclusion}
We have exhibited a simple model of  a limit order market in which order book events are described
in terms of a  Markovian queueing system. The analytical tractability of our model allows to compute various quantities of interest such as
\begin{itemize}
\item the distribution of the duration until the next  price change,
\item the distribution of price changes, and
\item the diffusion limit of the price process and its volatility.
\end{itemize}
in terms of parameters describing the order flow.
These results provide some insight into the relation between price dynamics and order flow in a limit order market.

We view this stylized model as a first step in the elaboration of
the analytical study of realistic stochastic models of order book
dynamics. Yet,    comparison with empirical data shows that even our
simple modeling set-up is capable of yielding useful analytical
insights into the relation between volatility and order flow, worthy
of being further pursued. Moreover, the  connection with
two-dimensional queueing systems
 allows  to use the rich analytical theory developed for these systems (see \cite{cohenboxma83}) to compute many other quantities. We hope to pursue further some of these ramifications in future work.

A relevant question is to examine which of the above results are robust to departures from the model assumptions
and whether the intuitions conveyed by our model  remain valid in a more general context where one or more of these assumptions are dropped.
This issue is further studied in a companion paper \cite{contlarrard2010b} where we explore a more general dynamic model relaxing some of the  assumptions above.


\ACKNOWLEDGMENT{The authors thank Jean-Philippe Bouchaud, Peter
Carr, Xin Guo, Charles Lehalle, Pete Kyle, Arseniy Kukanov, Costis
Maglaras and seminar participants at the European Meeting of
Statisticians (Athens, August 2010), Morgan Stanley, INFORMS 2010,
 the SIAM Conference on Financial Engineering and the Conference on Market microstructure (Paris, 2010) for helpful comments and    discussions.}

\end{document}